\documentclass{amsproc}
\begin{document}
\hskip 10cm CALT-68-2319

\hskip 10cm CITUSC/01-003
\vskip 1.3cm
\title[Matrix Theory and Noncommutative Gauge Theories]
{Energy-momentum Tensors in Matrix Theory\\
and in Noncommutative Gauge Theories}
\author[Y. Okawa]{Yuji Okawa}
\author[H. Ooguri]{Hirosi Ooguri}
\address{California Institute of Technology 452-48,
Pasadena, CA 91125, USA}
\email{okawa@theory.caltech.edu}
\email{ooguri@theory.caltech.edu}
\begin{abstract}
The energy-momentum tensor of Matrix Theory
is derived by computing disk amplitudes with 
one closed string and an arbitrary number of open
strings and by taking the DKPS limit. 
We clarify its relation to the energy-momentum 
tensor of the noncommutative gauge theory derived 
in our previous paper. 
\end{abstract}

\maketitle
\section{Introduction}
Noncommutative gauge theories can be realized by considering branes
in string theory with a strong NS-NS two-form field
\cite{Connes:1998cr}--\cite{Yin:1999ba}.
In our previous paper \cite{Okawa:2000sh}, we derived the energy-momentum
tensors of these theories by computing disk amplitudes with one closed 
string and an arbitrary number of open strings
and by taking the Seiberg-Witten 
limit \cite{Seiberg:1999vs} of these amplitudes.  
We found that the energy-momentum tensors involve
the open Wilson lines
\cite{Ishibashi:2000hs}--\cite{Das:2001ur}.\footnote
{Various aspects of the open Wilson lines and related issues
were discussed in \cite{Hyun:2000ma}--\cite{Okuyama:2001sw}.}
However, they do not reduce to
the ones in the commutative theories in the limit $\theta \rightarrow 0$,
where $\theta$ is the noncommutative parameter. This is because
the Seiberg-Witten limit does not commute with the commutative 
limit.\footnote{In this paper, we study the energy-momentum
tensor coupled to gravity in the bulk. The energy-momentum
tensor derived in the N\"other procedure, such as the one
discussed in \cite{Gerhold:2000ik}, would
couple to the open string metric on the branes.}
We also found that the energy-momentum 
tensors are conserved in interesting ways. 
In particular, in theories derived from bosonic string,
the energy-momentum tensors are kinematically 
conserved, namely, the conservation 
law holds identically for any field configuration
irrespective of the equations of motion. 

It turns out that we can use the same method
to derive the energy-momentum tensor
of Matrix Theory \cite{Banks:1997vh}.\footnote
{See also a recent review \cite{Taylor:2001vb}.}
If we perform the dimensional reduction of the
noncommutative theories along their noncommutative
directions, the Seiberg-Witten limit 
reduces to the DKPS limit \cite{Douglas:1997yp}
(or the Sen-Seiberg limit \cite{Sen:1998we, Seiberg:1997ad}
in the context of Matrix Theory
\cite{Banks:1997vh, Susskind:1997cw}).
Therefore, the energy-momentum tensor of Matrix Theory
can be derived as a special case of the computation 
in \cite{Okawa:2000sh} where there are no noncommutative 
directions. In \cite{Kabat:1998sa, Taylor:1999tv},
the energy-momentum tensor of Matrix Theory was 
deduced from computations of one-loop amplitudes 
in Matrix Theory and from their comparison with graviton exchange
amplitudes. The energy-momentum tensor we derive from
the disk amplitudes perfectly agrees with that obtained 
in \cite{Kabat:1998sa, Taylor:1999tv}
including the structure of the higher moments.

It has been shown
in \cite{deWit:1988ig, Banks:1997vh, Ishibashi:1997xs} and
\cite{Taylor:1997ik}--\cite{Seiberg:2000zk}
that noncommutative gauge theories are realized
as certain backgrounds of Matrix Theory and of the IIB matrix model.
This implies a natural correspondence of
gauge invariant observables
between Matrix Theory and noncommutative gauge theories.
In fact, it was in this context \cite{Ishibashi:2000hs}
that the open Wilson lines were discovered
as a basis for observables in noncommutative theories.
In \cite{Das:2001ur}, it was proposed to use this correspondence to derive the 
energy-momentum tensors of the noncommutative 
theories.\footnote{See also \cite{Liu:2001ps}.} In this paper, 
we carry out this proposal explicitly.
We find that the resulting
energy-momentum tensors agree exactly with those derived in
\cite{Okawa:2000sh},
both in the superstring case and in the bosonic string case.
The correspondence
between Matrix Theory and noncommutative theories also leads to 
a simpler proof of the conservation law of the energy-momentum tensors,
which came out rather miraculously in our earlier work \cite{Okawa:2000sh}.

This paper is organized as follows. In Section 2, 
we will derive the energy-momentum tensor of Matrix Theory 
by computing disk amplitudes
with one closed string and an arbitrary number of open
strings and by taking the DKPS limit. As we mentioned in the above,
this can be regarded as a special case of the computation
in \cite{Okawa:2000sh}.
In Section 3, we use the correspondence
between Matrix Theory and noncommutative 
gauge theories to derive the energy-momentum tensors in noncommutative 
theories and find an agreement with our earlier result.
We will study the case of bosonic string in Section 4.

\bigskip

Let us summarize the notations used in this paper.
The coordinates in the bulk are denoted by $x^M$.
For the discussion regarding the DKPS limit of D-branes,
we use the late Greeks ($\mu, \nu, ...$) for the directions
along the branes and $I, J, \ldots$ for those
transverse to the branes:
\begin{equation}
x^M = \begin{cases}
x^\mu & \text{for the directions along the branes}, \\
x^I & \text{for the transverse directions}.
\end{cases}
\end{equation}
In the context of noncommutative gauge theories,
we use the romans ($i, j, ...$)
for the noncommutative directions on the branes,
the late Greeks ($\mu, \nu, ...$) for the commutative directions,
and the early Greeks ($\alpha, \beta, ...$) for
the directions transverse to the branes:
\begin{equation}
x^M = \begin{cases}
x^\mu & \text{for the commutative directions
along the branes}, \\
x^i & \text{for the noncommutative directions
along the branes}, \\
x^\alpha & \text{for the transverse directions}.
\end{cases}
\end{equation}
As we will see, the transverse directions $x^I$
transmute into $x^i$ and $x^\alpha$.

In the noncommutative directions, the coordinates obey the
Heisenberg relation,
\begin{equation}
[x^i, x^j] = -i \theta^{ij}.
\end{equation}
The closed string metric is denoted by $g_{MN}$ and the
open string metric $G_{MN}$ in the zero-slope limit
is given by
\begin{equation}\label{metric}
G^{ij} = \frac{1}{(2\pi \alpha')^2}
\theta^{im} \theta^{jn} g_{mn}, \quad
G^{\mu\nu} = g^{\mu\nu}.
\end{equation}
The noncommutative gauge theory is defined by taking the zero-slope
limit, $\alpha' \rightarrow 0$, while keeping $G^{ij}$ and $\theta^{ij}$
finite (the Seiberg-Witten limit \cite{Seiberg:1999vs}).
This means that the closed string metric along the noncommutative
direction $g_{ij}$ is scaled as ${\alpha'}^2$.

\section{Energy-momentum tensor of Matrix Theory}

In \cite{Okawa:2000sh}, the energy-momentum tensors of noncommutative 
gauge theories are derived by computing the disk amplitudes
with one closed string and an arbitrary number of open strings.
It turns out that this method 
is applicable to the computation of the actions of multiple D-branes
in weakly curved background, or equivalently, the energy-momentum tensor
of Matrix Theory. 

The action of a single D$p$-brane in curved space is given by
\begin{equation}
\label{single-brane-action}
S = -\frac{1}{g_{YM}^2}
\int dx \sqrt{- \det g} \biggl[
\frac{1}{4} g^{\mu \nu} g^{\rho \sigma}
F_{\mu \rho} F_{\nu \sigma}
+ \frac{1}{2 (2 \pi \alpha')^2} g^{\mu \nu}
g_{IJ} \partial_\mu Y^I \partial_\nu Y^J
\biggr],
\end{equation}
where the scalar fields $Y^I (x)$ describe
the location of the brane in the transverse directions
and the Yang-Mills coupling constant $g_{YM}$
is given in terms of the string coupling constant $g_s$
and $\alpha'$ as follows:
\begin{equation}
g_{YM}^2 = g_s (2 \pi)^{p-2} {\alpha'}^{\frac{p-3}{2}}.
\end{equation}
The integral $\int dx$ is over the $p+1$ dimensional worldvolume
of the D-brane.
The action (\ref{single-brane-action}) is obtained
as a zero-slope limit $\alpha' \to 0$
known as the DKPS limit \cite{Douglas:1997yp}
(or the Sen-Seiberg limit \cite{Sen:1998we, Seiberg:1997ad}
in the context of Matrix Theory \cite{Banks:1997vh, Susskind:1997cw})
given by
\begin{equation}
\label{DKPS-limit}
Y^I \sim O(\alpha'), \quad k_I \sim O({\alpha'}^{-1}),
\end{equation}
while the other quantities
$A_\mu, k_\mu, g_{\mu \nu}, g_{IJ}$ and $g_{YM}$
are kept finite as $\alpha' \to 0$.
Here $k_M=(k_\mu, k_I)$ is the Fourier mode conjugate
to the coordinate $x^M=(x^\mu, x^I)$. In this limit,
$\Phi^I$ defined by
\begin{equation}
\label{Phi}
\Phi^I (x) = \frac{Y^I (x)}{2 \pi \alpha'},
\end{equation}
is kept finite. 
In the case of multiple D-branes, where $A_\mu$ and $Y^I$ become matrix
valued, a complete action in curved space is not known
although its various aspects have been studied in 
\cite{Kabat:1998sa, Taylor:1999tv}
and \cite{Douglas:1996sw}--\cite{Dasgupta:2000df}.
Here we will study the first term in
the expansion, 
\begin{equation}
g_{IJ} (x, Y(x)) = \eta_{IJ} + h_{IJ}
\exp \left[ ik_\mu x^\mu + ik_I Y^I(x) \right] + O(h^2).
\end{equation}
The DKPS limit (\ref{DKPS-limit}) defined in this
way is the same as the Seiberg-Witten limit of noncommutative
gauge theories as far as the commutative directions are concerned.
Therefore, we can derive the energy-momentum tensors of multiple D-branes
by considering 
the energy-momentum tensors of noncommutative gauge theories
found in \cite{Okawa:2000sh} and by taking
their dimensional reductions in the noncommutative directions.

It would be instructive to recapitulate the derivation
of the energy-momentum tensor in \cite{Okawa:2000sh},
in the language of Matrix Theory.
The worldsheet action is
\begin{equation}
S = \frac{1}{2 \pi} \int d^2 \sigma \left[
\frac{2}{\alpha'} g_{MN} \partial X^M \bar{\partial} X^N
+ \frac{1}{\alpha'} g_{MN} \psi^M \bar{\partial} \psi^N
+ \frac{1}{\alpha'} g_{MN} \bar{\psi}^M \partial \bar{\psi}^N
\right].
\end{equation}
When the worldsheet disk is represented by the upper-half plane
of the complex plane,
the propagators for $X^M (z)$ are
\begin{align}
& \langle X^M (z) X^N (w) \rangle \\ \notag
& = - \alpha' \biggl[ g^{MN} \log|z-w|
- g^{MN} \log|z-\bar{w}|
+ 2 G^{MN} \log|z-\bar{w}| \biggr],
\end{align}
and those for the fermions are
\begin{align}
\langle \psi^M (z) \psi^N (w) \rangle
&= \frac{\alpha'}{z-w} g^{MN}, \\
\notag \langle \psi^M (z) \bar{\psi}^N (\bar{w}) \rangle
&= \frac{\alpha'}{z-\bar{w}} \left(
-g^{MN} + 2G^{MN} \right), \\
\notag \langle \bar{\psi}^M (\bar{z}) \psi^N (w) \rangle
&= \frac{\alpha'}{\bar{z} - w}
\left( -g^{MN} + 2G^{MN} \right), \\
\notag \langle \bar{\psi}^M (\bar{z}) \bar{\psi}^N (\bar{w}) \rangle
&= \frac{\alpha'}{\bar{z} - \bar{w}} g^{MN},
\end{align}
where
\begin{equation}
G^{\mu \nu} = g^{\mu \nu}, \quad
G^{IJ} = 0.
\end{equation}
The fermions on the boundary $\Psi^M (t)$ are defined by
\begin{equation}
\Psi^\mu (t) = \frac{1}{2}
\left[ \psi^\mu (t) + \bar{\psi}^\mu (t) \right], \quad
\Psi^I = \frac{1}{2}
\left[ \psi^I (t) - \bar{\psi}^I (t) \right],
\end{equation}
and the boundary-boundary and bulk-boundary propagators
for the fermions are given by
\begin{align}
\langle \Psi^\mu (t) \Psi^\nu (t') \rangle
&= \frac{\alpha'}{t - t'} g^{\mu \nu}, &
\langle \Psi^I (t) \Psi^J (t') \rangle
&= \frac{\alpha'}{t - t'} g^{IJ}, \\
\langle \psi^\mu (z) \Psi^\nu (t) \rangle
&= \frac{\alpha'}{z - t} g^{\mu \nu}, &
\langle \bar{\psi}^\mu (\bar{z}) \Psi^\nu (t) \rangle
&= \frac{\alpha'}{\bar{z} - t} g^{\mu \nu}, \\
\notag
\langle \psi^I (z) \Psi^J (t) \rangle
&= \frac{\alpha'}{z - t} g^{IJ}, &
\langle \bar{\psi}^I (\bar{z}) \Psi^J (t) \rangle
&= -\frac{\alpha'}{\bar{z} - t} g^{IJ},
\end{align}
where $t$ and $t'$ are on the boundary.

It is convenient to use the vertex operator in the $(-1,-1)$-picture
for the closed string,
\begin{equation}
V^{(-1,-1)}(z) = \frac{1}{2} \delta(\gamma) \delta(\bar{\gamma})
h_{MN} (k) \psi^M (z) \bar{\psi}^N (\bar{z}) e^{ikX(z)},
\end{equation}
where $\gamma$ and $\bar{\gamma}$ are bosonic ghosts,
and the operator in the $0$-picture for open string,
\begin{multline}
\label{U^{(0)}}
U^{(0)} (t)
= A_\mu (X) \frac{d X^\mu}{dt}
+ \Phi^I (X) i g_{IJ} \partial_\perp X^J
- F_{\mu \nu} (X) \Psi^\mu \Psi^\nu \\
-2 D_\mu \Phi^I (X) g_{IJ} \Psi^\mu \Psi^J
-i g_{IK} g_{JL} [\Phi^I (X), \Phi^J (X)] \Psi^K \Psi^L.
\end{multline}
Here we used $\Phi^I$ defined in (\ref{Phi}) instead of $Y^I$
to make it clear
that the open string vertex operator is $O(1)$
in the zero-slope limit.
The correlation function which we need to evaluate is then
\begin{equation}
(z-\bar{z})^2 \langle
V^{(-1,-1)} (z)~
tr P \exp \left( i \int_{-\infty}^\infty dt~
U^{(0)} (t) \right) \rangle,
\end{equation}
with a proper gauge-fixing of the $SL(2,R)$ invariance.
Here the symbol $P$ denotes the path-ordering
with respect to $t$.

By definition, 
all the propagators are $O(\alpha')$
whereas the open vertex operator is $O(1)$
in the zero-slope limit.
Therefore, it costs $\alpha'$ each time we use a propagator.
However, when we make a contraction between
an open string vertex operator and
the $e^{i k_I X^I}$ part of the graviton vertex operator, 
the factor $\alpha'$ is cancelled
by the fact that the momentum $k_I$ scales as $O({\alpha'}^{-1})$,
so the contraction remains finite for any number of 
open string vertex operators. 
\begin{align}
& \langle e^{ikX (z)}
tr \prod_a \Phi^I (X(t_a)) i g_{IJ} \partial_\perp X^J (t_a)
\rangle \\ \notag
& = \int dx~ e^{i k_\mu x^\mu}~
tr \prod_a 2 \pi \alpha' k_I \Phi^I (x)
\frac{\partial \tau (t_a, z)}{\partial t_a}.
\end{align}
Here the function $\tau (t, z)$ is given by
\begin{equation}
\tau (t, z) = \frac{1}{2 \pi i}
\log \left( \frac{t-z}{t-\bar{z}} \right),
\end{equation}
and we used that
\begin{equation}
i \partial_\perp \log |z-t|^2
= \frac{z-\bar{z}}{(z-t)(\bar{z}-t)}
= 2 \pi i \frac{\partial \tau (t, z)}{\partial t}.
\end{equation}
For a fixed value of $z$,
$\tau (t,z)$ is a monotonically increasing function of $t$ and
\begin{equation}
\tau (\infty, z) - \tau (-\infty, z) =1.
\end{equation}
Therefore, we have
\begin{align}
& \langle e^{ikX (z)}
~tr P \exp \left( i \int_{-\infty}^\infty dt~
\Phi^I (X(t)) i g_{IJ} \partial_\perp X^J (t) \right)
\rangle \\ \notag
& = \int dx~ e^{i k_\mu x^\mu}~
tr~P \exp \left( i \int_0^1 d \tau~
2 \pi \alpha' k_I \Phi^I (x) \right).
\end{align}
Note that the path-ordering with respect to $\tau$
on the right-hand side is inherited from that
with respect to $t$ on the left-hand side
because $\tau$ is a monotonically
increasing function of $t$.
Note also that the $A_\mu (X) d X^\mu /dt$ part of
the open string vertex operator (\ref{U^{(0)}})
does not contribute to the energy-momentum tensor
in the zero-slope limit
because there is no divergent factor like $k_I$
which can compensate the factor of $\alpha'$
coming from the propagator.

We also need to contract the fermions
in the closed string vertex operator
to obtain a non-vanishing result.
The contribution to the graviton vertex operator
starts at $O({\alpha'}^3)$ with three contractions of the fermions.
\begin{align}
& (z-\bar{z}) \langle \frac{1}{2} \left[
\psi^I (z) \bar{\psi}^J (\bar{z})
+ \psi^J (z) \bar{\psi}^I (\bar{z}) \right]
tr P \exp \left( i \int_{-\infty}^\infty dt~
U^{(0)} (t) \right) \rangle \\
\notag &= 8 \pi^2 {\alpha'}^3 ~tr \biggl[
g^{\mu \nu} \int_0^1 d \tau_1 D_\mu \Phi^I (x)
\int_0^1 d \tau_2 D_\nu \Phi^J (x) \\
\notag & \qquad + g_{KL} \int_0^1 d \tau_1 [\Phi^I (x), \Phi^K (x)]
\int_0^1 d \tau_2 [\Phi^L (x), \Phi^J (x)] \biggr]
+ O({\alpha'}^4) \\
\notag & \qquad +~ \text{the dilaton part}.
\end{align}
After taking into account the contributions from the ghosts,
we obtain the components $T^{IJ}$
of the energy-momentum tensor
in the case of superstring as follows:
\begin{multline}
2 \alpha' \int dx~ e^{ikx}~ tr \biggl[ \exp \left(
i \int_0^1 d \tau~
k_I Y^I (x) \right) \\
\times \Bigl[
g^{\mu \nu} \int_0^1 d \tau_1 D_\mu Y^I (x)
\int_0^1 d \tau_2 D_\nu Y^J (x) \\
+ \frac{g_{KL}}{(2 \pi \alpha')^2}
\int_0^1 d \tau_1 [Y^I (x), Y^K (x)]
\int_0^1 d \tau_2 [Y^L (x), Y^J (x)] \Bigr] \biggr],
\end{multline}
where we used $Y^I$ instead of $\Phi^I$ in (\ref{Phi}).
Note that the ordering of the fields is determined
by the values of $\tau$, $\tau_1$ and $\tau_2$
which are inherited from the positions of
open string vertex operators.
We can transform the expression for $T^{IJ}$
into a more convenient form.
For any operators $A$ and $B$,
\begin{align}
& tr \left[
\exp \left( i \int_0^1 d \tau~ kY \right)
\int_0^1 d \tau_1~ A
\int_0^1 d \tau_2~ B
\right] \\
\notag &
= \int_0^1 d \tau_1 \int_{\tau_1}^1 d \tau_2~ tr~
A~ e^{i (\tau_2 - \tau_1) kY}
B~ e^{i (1 + \tau_1 - \tau_2) kY} \\
\notag &
\quad + \int_0^1 d \tau_1 \int_0^{\tau_1} d \tau_2~ tr~
A~ e^{i (1 - \tau_1 + \tau_2) kY}
B~ e^{i (\tau_1 - \tau_2) kY} \\
\notag &
= \int_0^1 d \tau \left[
\int_0^{1 - \tau} d \tau'
+ \int_{1 - \tau}^1 d \tau' \right]
tr~ A~ e^{i \tau kY} B~ e^{i (1 - \tau) kY} \\
\notag &
= \int_0^1 d \tau~ tr~
A~ e^{i \tau kY} B~ e^{i (1 - \tau) kY}.
\end{align}
Therefore, we can write
\begin{multline}
T^{IJ}
= 2 \alpha' \int dx~ e^{ikx} ~tr \int_0^1 d \tau~
\biggl[
g^{\mu \nu} D_\mu Y^I e^{i \tau k_I Y^I}
D_\nu Y^J e^{i (1-\tau) k_I Y^I} \\
+ \frac{g_{KL}}{(2 \pi \alpha')^2}
[Y^I, Y^K] e^{i \tau k_I Y^I}
[Y^L, Y^J] e^{i (1-\tau) k_I Y^I}
\biggr].
\end{multline}
Furthermore, we can carry out the remaining $\tau$-integral
and the result is expressed in terms of
the symmetrized trace \cite{Tseytlin:1997cs}.\footnote{
We thank S. Das, W. Taylor and S. Trivedi for discussion
on this point.} To see this we note that,
for any operators $A$ and $B$, 
\begin{align}
& \quad \int_0^1 d \tau~ tr
\left[ A e^{i\tau k Y} B e^{i(1-\tau)kY} \right] \\
&= \sum_{n, m=0}^\infty
\int_0^1 d\tau~ tr \left[ A \frac{1}{n!}(i \tau k Y)^n
B \frac{1}{m!} (i(1-\tau)kY)^m \right] \notag \\
&= \sum_{n,m=0}^\infty \frac{n! m! }{(n+m+1)!}
tr \left[ A \frac{1}{n!} (ikY)^n
B \frac{1}{m!} (ikY)^m \right] \notag \\
&= \sum_{p=0}^\infty \frac{1}{p!} \frac{1}{p+1}
\sum_{q=0}^p 
tr \left[ A (ikY)^q B (ikY)^{p-q} \right] .\notag \end{align}
Since
\begin{align}
Str\left( C^p A B\right)
&= \frac{1}{(p+1)!}
\sum_{q=0}^p p! ~tr \left[ A C^q B C^{p-q} \right]
\\ \notag
&= \frac{1}{p+1} \sum_{q=0}^p tr\left[ A C^q B C^{p-q} \right],
\end{align}
where $Str$ is the symmetrized trace, we can write
\begin{equation}
\quad \int_0^1 d \tau~ tr
\left[ A e^{i\tau k Y} B e^{i(1-\tau)kY} \right] 
= Str\left( e^{ikY} AB\right).
\end{equation}
Thus $T^{IJ}$ is expressed as
\begin{multline}
T^{IJ}
= 2 \alpha' \int dx~ e^{ikx}~ Str 
\biggl[
e^{i k_I Y^I} g^{\mu \nu} D_\mu Y^I D_\nu Y^J \\
+ e^{i k_I Y^I} \frac{g_{KL}}{(2 \pi \alpha')^2}
[Y^I, Y^K] [Y^L, Y^J]
\biggr],
\end{multline}
where we symmetrized over all possible orderings of
$Y^I$, $D_\mu Y^I$ and $[Y^I, Y^J]$ inside the trace.
Similarly, we can derive the expressions for
$T^{\mu J}$ and $T^{\mu \nu}$.
\begin{align}
T^{\mu J}
&= 2 \alpha' g^{\mu \mu'} \int dx~ e^{ikx}~
tr \left[ e^{i k_I Y^I} D_{\mu'} Y^J \right], \\
T^{\mu \nu}
&= 2 \alpha' g^{\mu \nu} \int dx~ e^{ikx}~
tr \left[ e^{i k_I Y^I} \right].
\end{align}
These expressions agree with those derived in
\cite{Kabat:1998sa, Taylor:1999tv, Taylor:1999gq, Taylor:2000pr}
using one-loop amplitudes of Matrix Theory.
Our result for the graviton can be extended
to all other closed string states including massive ones
as discussed in \cite{Okawa:2000sh}.

The appearance of the symmetrized trace is not an obvious consequence
of string theory computation. To the contrary, in the case
of bosonic string,
the ordering of operators in the energy-momentum tensor
does not obey the symmetrized trace prescription as we will see
in Section 4.

The conservation
of the energy-momentum tensor,
\begin{equation}
k_\mu T^{\mu J} + k_I T^{IJ} = 0,
\end{equation}
has been verified by Van Raamsdonk \cite{VanRaamsdonk:1999in}. 
For the D$(-1)$-branes, it goes as
\begin{align}
k_I T^{IJ} & \propto
tr \int_0^1 d \tau~ [k \cdot Y, Y^K]
e^{i \tau k \cdot Y} [Y_K, Y^J] e^{i (1-\tau) k \cdot Y} \\
&= i~ tr \int_0^1 d \tau~ \frac{d}{d \tau} \left(
Y^K e^{i \tau k \cdot Y} [Y_K, Y^J] e^{i (1-\tau) k \cdot Y}
\right) \notag \\
&= -i~ tr [Y^K, [Y_K, Y^J]] e^{i k \cdot Y} \notag \\
&=0 , \notag
\end{align}
where $k \cdot Y \equiv k_I Y^I$ and $Y_K \equiv g_{KL} Y^L$.
In the last line, we used the equation of motion of Matrix Theory,
\begin{equation} [Y^K, [ Y_K, Y^J]]=0. \end{equation}
It is easily generalized to the case of D$p$-branes
using the identity
\begin{equation}
D_\mu (e^{A}) = \int_0^1 d \tau~
e^{(1-\tau) A} D_\mu A~ e^{\tau A}.
\end{equation}

\section{Relation to noncommutative gauge theory}

Given the energy-momentum tensor of Matrix Theory,
we can derive those of noncommutative gauge theories
following the suggestion in \cite{Das:2001ur}.
Let us carry this out explicitly here. 
We divide the transverse directions $x^I$
into two, $x^I =(x^\alpha, x^i)$. To fit with
the standard notation in noncommutative gauge
theories, we rescale the coordinates $x^i$
so that $g_{ij} \sim O({\alpha'}^2)$. 
On the other hand, $g_{\alpha\beta}$ as well
as $g_{\mu\nu}$ remains finite. 
Consider the background
\begin{equation}
\label{background}
Y^\alpha =0, \quad Y^i = x^i \quad 
\text{such that} \quad
[x^i, x^j] = -i \theta^{ij},
\end{equation}
and expand the scalar fields around this background as follows:
\begin{equation}
\label{expansion}
Y^\alpha = 2 \pi \alpha' \Phi^\alpha \sim O(\alpha'), \quad
Y^i = x^i + \theta^{ij} A_j \sim O(1).
\end{equation}
We take the limit $k_\alpha \sim O({\alpha'}^{-1})$
and $g_{ij} \sim O({\alpha'}^2)$ keeping 
$(\Phi^\alpha, A^i, A_\mu)$, $(k_i, k_\mu)$
and $(g_{\alpha\beta}, G_{\mu\nu} = g_{\mu\nu})$ finite. 
This is identical to the limit taken in \cite{Okawa:2000sh}
following \cite{Seiberg:1999vs}.
An open Wilson line, which constitutes a basic ingredient
of the energy-momentum tensors
of the noncommutative gauge theories,
emerges in this background as follows \cite{Ishibashi:2000hs}:
\begin{multline}
tr \exp \left[ i k_I Y^I \right]
= tr \exp
\left[ i k_i x^i + i l^i A_i + i y_\alpha \Phi^\alpha \right] \\
\propto \int dx~ \ast \biggl[ \exp \left(
i \int_0^1 d \tau \left( l^i A_i (x + l \tau)
+ y_\alpha \Phi^\alpha (x + l \tau) \right) \right)
e^{i k_i x^i} \biggr],
\end{multline}
where $l^i=k_j \theta^{ji}$, $y_\alpha = 2\pi \alpha'k_\alpha$.
Here we mapped the matrices
into the functions on noncommutative space with the star product
on the right-hand side.
The symbol $*[ \ldots ]$ means that we take the star product
in the expression in $[ \ldots ]$
taking into account the path-ordering along the open Wilson line
between $x^i$ and $x^i+l^i$.
The integral $\int dx$ here
is over the noncommutative directions $x^i$ along the branes
which constitutes the integral over the whole worldvolume
of the branes together with the previous integral over
the commutative directions $x^\mu$.
In the zero-slope limit,
the on-shell condition of the graviton is
\begin{equation}
G_{ij} l^i l^j + g^{\alpha\beta} y_\alpha y_\beta = 0.
\end{equation}

For the cases with operator insertions along the Wilson line,
the following formulas are useful:
\begin{align}
D_\mu Y^I &= \begin{cases}
2 \pi \alpha' D_\mu \Phi^\alpha
& \text{when $I=\alpha$}, \\
\theta^{ij} F_{\mu j}
& \text{when $I=i$},
\end{cases}\\
[Y^I, Y^J] &= \begin{cases}
(2 \pi \alpha')^2 [\Phi^\alpha, \Phi^\beta]
& \text{when $(I, J)=(\alpha, \beta)$}, \\
2 \pi i \alpha' \theta^{jj'} D_{j'} \Phi^\alpha
& \text{when $(I, J)=(\alpha, j)$}, \\
-i \theta^{ii'} \theta^{jj'} ( F_{i'j'}-\theta^{-1}_{i'j'})
& \text{when $(I, J)=(i, j)$},
\end{cases}
\end{align}
and
\begin{multline}
tr \int_0^1 d \tau~ A e^{i \tau k_I Y^I}
B e^{i (1-\tau) k_I Y^I} \\
\propto \int dx~ \ast \biggl[ \exp \left(
i \int_0^1 d \tau \left( l^i A_i (x + l \tau)
+ y_\alpha \Phi^\alpha (x + l \tau) \right) \right) \\
\times \int_0^1 d \tau_1~ A ( x + l \tau_1 )
\int_0^1 d \tau_2~ B ( x + l \tau_2 )
~e^{i k_i x^i} \biggr].
\end{multline}
It is straightforward to verify that all the components of
energy-momentum tensor of the noncommutative gauge theory
in the case of superstring derived in \cite{Okawa:2000sh}
are correctly reproduced
including the relative coefficients.
For example, the expression for $T^{ij}$ is given by
\begin{multline}
\label{T^{ij}}
T^{ij}
= 2 \alpha' \theta^{i i'} \theta^{j j'} \\
 \times \int dx \ast \biggl[ e^{ikx} \exp \left(
i \int_0^1 d \tau \left( l^i A_i (x + l \tau)
+ y_\alpha \Phi^\alpha (x + l \tau) \right) \right) \\
\times \int_0^1 d \tau_1 \int_0^1 d \tau_2
\Bigl[ G^{mn}
\bigl( F_{i' m} (x + l \tau_1) -\theta^{-1}_{i' m} \bigr)
\bigl( F_{j' n} (x + l \tau_2) -\theta^{-1}_{j' n} \bigr) \\
+ G^{\mu \nu} F_{i' \mu} (x + l \tau_1) F_{j' \nu} (x + l \tau_2)
+ g_{\alpha \beta} D_{i'} \Phi^\alpha (x + l \tau_1)
D_{j'} \Phi^\beta (x + l \tau_2) \Bigr] \biggr],
\end{multline}
and the other components are presented in the Appendix.

Two important checks have been made in \cite{Okawa:2000sh}
to demonstrate that this is indeed the correct expression
for the energy-momentum tensor.
One is the consistency with the way the action of
the noncommutative gauge theory \cite{Seiberg:1999vs}
depends on the bulk metric $g_{MN}$.
\begin{multline}
\label{action}
S= -\frac{1}{g_{YM}^2}\int dx
\sqrt{-\det G} \ast \biggl[
  ~\frac{1}{4} G^{MN} G^{PQ} (F_{MP}-\theta^{-1}_{MP})
(F_{NQ}-\theta^{-1}_{NQ}) \\
+ \frac{1}{2} G^{MN} g_{\alpha\beta}
D_M \Phi^\alpha  D_N \Phi^\beta
- \frac{1}{4} g_{\alpha\beta}g_{\gamma\delta}
 [\Phi^{\alpha}, \Phi^{\gamma}]
 [\Phi^{\beta}, \Phi^\delta] \biggr] .
\end{multline}
Here $\theta^{-1}_{MN}=0$ unless $(M,N)=(i,j)$.
Since the action is derived assuming that the bulk metric is flat,
we can compare the variation $\partial S/\partial g_{MN}$ with
the zero-momentum limit of the energy-momentum tensor.

The other one is the conservation of the energy-momentum
tensor,
\begin{equation}
\label{conservation}
k_M T^{MN} =0.
\end{equation}
We now have a simpler proof of this
in the language of the matrix model
discussed in the preceding section.

One may have expected that the energy-momentum tensor of the noncommutative
theory should reduce to that of the commutative theory (theory defined on
the commutative space) in the limit $\theta^{ij}\rightarrow 0$.
\begin{equation}
\label{commutative}
T_{comm}^{ij} = G^{ii'} G^{jj'} (F_{i'm}F_{j'n} G^{mn}
 -\frac{1}{4} G_{i'j'} F^2 + \cdots ), ~~~ \text{etc}.
\end{equation}
Here $\cdots$ denotes terms containing the scalar field
$\Phi^\alpha$ $etc$.
It turned out that they are different in the 
following three accounts:
\begin{enumerate}
\item
In the commutative case (\ref{commutative}), 
the metric $G^{ii'}G^{jj'}$ is used 
to raise
the 
indices of the energy-momentum tensor, whereas 
in the 
noncommutative case
(\ref{T^{ij}}) 
we use $\theta^{ii'}\theta^{jj'}$
in the corresponding 
term.
\item
In the noncommutative case, the field strength $F_{ij}$ is shifted by
$\theta^{-1}_{ij}$. 
\item 
In the noncommutative case, there are no terms that correspond to 
the term $-\frac{1}{4} G^{ij}F^2$ 
in (\ref{commutative}).
\end{enumerate}
All these differences are important in order to maintain
the consistency with the metric dependence of the
Seiberg-Witten action (\ref{action})
and the conservation of the energy-momentum tensor.

\section{Bosonic string}
We can use the same technique to 
derive the energy-momentum tensor of multiple D-branes
in the bosonic string theory in the zero-slope limit.
The closed string vertex operator is
\begin{equation}
V(z) = h_{MN} (k) \partial X^M \bar{\partial} X^N
e^{ikX(z)},
\end{equation}
and the open string vertex operator is nothing but
the bosonic part of $U^{(0)} (t)$ in (\ref{U^{(0)}}).
In the superstring case,
the contribution to the graviton coupling starts
at $O({\alpha'}^3)$.
In the bosonic case, on the other hand, it starts
at $O({\alpha'}^2)$ coming from the product of
the following contractions:
\begin{align}
\langle \partial X^I (z) i \partial_\perp X^K (t) \rangle
&= -\frac{\alpha' g^{IK}}{(z-t)^2}
= -\frac{2 \pi i \alpha' g^{IK}}{z-\bar{z}}
e^{-2 \pi i \tau (t,z)} \frac{\partial \tau (t,z)}{\partial t},
\\ \notag
\langle \bar{\partial} X^J (\bar{z})
i \partial_\perp X^L (t) \rangle
&= \frac{\alpha' g^{JL}}{(\bar{z}-t)^2}
= \frac{2 \pi i \alpha' g^{JL}}{z-\bar{z}}
e^{2 \pi i \tau (t,z)} \frac{\partial \tau (t,z)}{\partial t}.
\end{align}
The expression for $T^{IJ}$ is then
\begin{align}
\label{bosonic-T^{IJ}}
T^{IJ}
&= - \int dx~ e^{ikx}~ tr \Biggl[ \exp \left(
i \int_0^1 d \tau~
k_I Y^I (x) \right) \\
&\qquad \times \biggl[
\int_0^1 d \tau_1~ e^{-2 \pi i \tau_1} Y^I (x)
\int_0^1 d \tau_2~ e^{ 2 \pi i \tau_2} Y^J (x)
+ (I \leftrightarrow J) \biggr] \Biggr] \notag \\
&= - \int dx~ e^{ikx}~ tr \int_0^1 d \tau~
\biggl[
e^{2 \pi i \tau} Y^I e^{i \tau k_I Y^I}
Y^J e^{i (1-\tau) k_I Y^I} + (I \leftrightarrow J)
\biggr]. \notag
\end{align}
It is interesting to note
that this leading term in the zero-slope limit
vanishes in the case of a single brane since
all the scalar fields commute and the only $\tau$
dependence of the integrand is $e^{2\pi i\tau}$.
Thus $\int_0^1 d\tau 
e^{2\pi i \tau} = 0$.
Another consequence of
the existence of the factor $e^{2 \pi i \tau}$
is that the ordering of the scalar fields no longer
obeys the symmetrized trace prescription,
\begin{equation}
T^{IJ} \ne -2 \int dx~ e^{ikx}~ Str \biggl[
e^{i k_I Y^I} Y^I Y^J \biggr].
\end{equation}
One can see this easily by considering the
case of a single D-brane where $T^{IJ}=0$ in this limit
as we just mentioned 
whereas the symmetrized trace on the right-hand side 
is non-vanishing.

To see the relation of this energy-momentum
tensor to those of the noncommutative gauge theories
derived in \cite{Okawa:2000sh}
(reproduced in the Appendix of this paper),
the following identity is useful, which
can be derived by integration by parts,
\begin{align}
& 
tr \int_0^1 d \tau~ e^{\pm 2 \pi i \tau} A ~e^{i \tau k \cdot Y}
B ~e^{i (1-\tau) k \cdot Y} \\
&= tr \int_0^1 d \tau~ \frac{1}{\pm 2 \pi i} \frac{d}{d \tau}
\left( e^{\pm 2 \pi i \tau} \right)
A ~e^{i \tau k \cdot Y} B ~e^{i (1-\tau) k \cdot Y} \notag \\
&= - \frac{1}{\pm 2 \pi i}~ tr~ [A, B] ~e^{i k \cdot Y} \notag \\
& \qquad - \frac{1}{\pm 2 \pi i}
~tr \int_0^1 d \tau~ e^{\pm 2 \pi i \tau}
A ~e^{i \tau k \cdot Y} [i k \cdot Y, B] ~e^{i (1-\tau) k \cdot Y}\notag\\
&= - \frac{1}{\pm 2 \pi i}~ tr~ [A, B] ~e^{i k \cdot Y} \notag \\
& \qquad - \frac{1}{(2 \pi)^2}~
tr~ [A, [i k \cdot Y, B]] ~e^{i k \cdot Y} \notag \\
& \qquad + \frac{1}{(2 \pi)^2}
~tr \int_0^1 d \tau~ e^{\pm 2 \pi i \tau}
[i k \cdot Y, A] ~e^{i \tau k \cdot Y}
[i k \cdot Y, B] ~e^{i (1-\tau) k \cdot Y} .\notag
\end{align}
After we symmetrize with respect to $A$ and $B$, 
the first term on the right-hand side vanishes. 
Combining this with 
\begin{align}
&[i k_I Y^I, x^j + \theta^{jm} A_{m}]
= - \theta^{jm} \left(
l^n (F_{mn} - \theta^{-1}_{mn}) + y_\beta D_m \Phi^\beta \right),\notag\\
&[x^i + \theta^{im} A_m, \mathcal{O}]
= -i \theta^{im} D_m \mathcal{O},\notag
\end{align}
it is then straightforward to see that 
(\ref{bosonic-T^{IJ}}) reproduces the energy-momentum tensors of the
noncommutative theories (\ref{boseij})--(\ref{boseab}).

The proof of the conservation of the energy-momentum tensor
$k_I T^{IJ} =0$ 
in the matrix form (\ref{bosonic-T^{IJ}}) of $T^{IJ}$ 
is given by 
\begin{align}
&tr \int_0^1 d \tau~ e^{\pm 2 \pi i \tau} k \cdot Y
~e^{i \tau k \cdot Y} Y^J ~e^{i (1-\tau) k \cdot Y} \\
&= tr \int_0^1 d \tau~ e^{\pm 2 \pi i \tau}
~e^{i \tau k \cdot Y} k \cdot Y~ Y^J ~e^{i (1-\tau) k \cdot Y}
\notag \\
&= \int_0^1 d \tau~ e^{\pm 2 \pi i \tau}
~tr~ k \cdot Y~ Y^J ~e^{i k \cdot Y} \notag \\
&= 0 \quad \text{because of the $\tau$ integration}. \notag
\end{align}
Note that the conservation holds without using 
the equation of motion. 
This proof is significantly simpler than the one given
in \cite{Okawa:2000sh}. 

\section*{Acknowledgments}

We would like to thank Sumit Das, Satoshi Iso, Washington Taylor
and Sandip Trivedi for discussions.
H. O. thanks the organizers of the conference, {\it Strings 2001},
at Tata Institute of Fundamental Research, Mumbai, India
for their hospitality and for giving him the opportunity to
present our work at the conference. 
 
The research was supported in part by
the DOE grant DE-FG03-92ER40701
and the Caltech Discovery Fund.

\appendix
\section{Energy-momentum tensors \\of noncommutative gauge theories}
\subsection{Superstring}
\begin{multline}
T^{ij}
= 2 \alpha' \theta^{i i'} \theta^{j j'} \\
 \times \int dx \ast \biggl[ e^{ikx} \exp \left(
i \int_0^1 d \tau \left( l^i A_i (x + l \tau)
+ y_\alpha \Phi^\alpha (x + l \tau) \right) \right) \\
\times \int_0^1 d \tau_1 \int_0^1 d \tau_2
\Bigl[ G^{mn}
\bigl( F_{i' m} (x + l \tau_1) -\theta^{-1}_{i' m} \bigr)
\bigl( F_{j' n} (x + l \tau_2) -\theta^{-1}_{j' n} \bigr) \\
+ G^{\mu \nu} F_{i' \mu} (x + l \tau_1) F_{j' \nu} (x + l \tau_2)
+ g_{\alpha \beta} D_{i'} \Phi^\alpha (x + l \tau_1)
D_{j'} \Phi^\beta (x + l \tau_2) \Bigr] \biggr],
\end{multline}
\begin{multline}
T^{\mu j}
= 2 \alpha' G^{\mu \mu'} \theta^{j j'} \\
\times \int dx \ast \biggl[ e^{ikx} \exp \left(
i \int_0^1 d \tau \left( l^i A_i (x + l \tau)
+ y_\alpha \Phi^\alpha (x + l \tau) \right) \right) \\
\times \int_0^1 d \tau' F_{\mu' j'} (x + l \tau') \biggr],
\end{multline}
\begin{multline}
T^{\alpha j}
= -4 \pi {\alpha'}^2 \theta^{j j'} \\
\times \int dx \ast \biggl[ e^{ikx} \exp \left(
i \int_0^1 d \tau \left( l^i A_i (x + l \tau)
+ y_\alpha \Phi^\alpha (x + l \tau) \right) \right) \\
\times \int_0^1 d \tau_1 \int_0^1 d \tau_2
\Bigl[ G^{mn}
D_{m} \Phi^\alpha (x + l \tau_1)
\bigl( F_{j' n} (x + l \tau_2) -\theta^{-1}_{j' n} \bigr) \\
\qquad \qquad \qquad \qquad
+ G^{\mu \nu} D_{\mu} \Phi^\alpha (x + l \tau_1)
F_{j' \nu} (x + l \tau_2) \\
+ g_{\beta \gamma} i [\Phi^\beta, \Phi^\alpha] (x + l \tau_1)
D_{j'} \Phi^\gamma (x + l \tau_2) \Bigr] \biggr],
\end{multline}
\begin{multline}
T^{\mu \nu}
= 2 \alpha' G^{\mu \nu} \\
\times \int dx \ast \biggl[ e^{ikx} \exp \left(
i \int_0^1 d \tau \left( l^i A_i (x + l \tau)
+ y_\alpha \Phi^\alpha (x + l \tau) \right) \right) \biggr],
\end{multline}
\begin{multline}
T^{\alpha \nu}
= 4 \pi {\alpha'}^2 G^{\nu \nu'} \\
\times \int dx \ast \biggl[ e^{ikx} \exp \left(
i \int_0^1 d \tau \left( l^i A_i (x + l \tau)
+ y_\alpha \Phi^\alpha (x + l \tau) \right) \right) \\
\times \int_0^1 d \tau'
D_{\nu'} \Phi^\alpha (x + l \tau') \biggr],
\end{multline}
\begin{multline}
T^{\alpha \beta}
= 8 \pi^2 {\alpha'}^3 \\
\times \int dx \ast \biggl[ e^{ikx} \exp \left(
i \int_0^1 d \tau \left( l^i A_i (x + l \tau)
+ y_\alpha \Phi^\alpha (x + l \tau) \right) \right) \\
\times \int_0^1 d \tau_1 \int_0^1 d \tau_2
\Bigl[ G^{ij}
D_i \Phi^\alpha (x + l \tau_1) D_j \Phi^\beta (x + l \tau_2) \\
\qquad \qquad \qquad \qquad \qquad \qquad
+ G^{\mu \nu} D_\mu \Phi^\alpha (x + l \tau_1)
D_\nu \Phi^\beta (x + l \tau_2) \\
- g_{\gamma \delta} [\Phi^\gamma, \Phi^\alpha] (x + l \tau_1)
[\Phi^\delta, \Phi^\beta] (x + l \tau_2) \Bigr] \biggr].
\end{multline}
\subsection{Bosonic string}
\begin{multline}\label{boseij}
T^{ij} =
\frac{\theta^{ii'}\theta^{jj'} +\theta^{ji'}\theta^{ij'}}
{(2\pi)^2} \\
\times \int dx \ast \left[e^{ikx} 
\exp \left( i\int_0^1 d\tau ~ l^i
A_i(x+l\tau)  + y_\alpha \Phi^\alpha(x+l \tau) \right)
\right. \\ \times \left\{
i\int_0^1 d\tau_1~ e^{-2\pi i \tau_1}\left( l^{m} F_{i'm}(x+l\tau_1)
+ y_\alpha D_{i'} \Phi^\alpha(x+l\tau_1) \right)
\right. \\ \times
i\int_0^1 d\tau_2~ e^{2\pi i \tau_2}\left( l^{n} F_{j'n}(x+l\tau_2)
+ y_\beta D_{j'} \Phi^\beta(x+l\tau_2) \right)
\\ \left.\left. + ~
i\int_0^1 d\tau' \Big( l^{m} D_{i'} F_{j'm}(x+l\tau') + y_\alpha D_{i'}
D_{j'} \Phi^\alpha(x+l\tau') \Big)\right\}\right].
\end{multline}
\begin{multline}\label{boseaj}
T^{\alpha j}
= \alpha' \theta^{jm}
\int dx~* \left[e^{ikx}~
\exp \left( i\int_0^1 d\tau ~ l^i
A_i(x+l\tau) + y_\alpha \Phi^\alpha(x+l \tau) \right)
\right. \\ \times
\left\{
\int_0^1 d\tau_1~ e^{-2\pi i \tau_1}  \Phi^\alpha(x+l\tau_1)
\right. \qquad \qquad \qquad \\
\qquad \qquad \qquad \times \left.
i \int_0^1 d\tau_2~ e^{2\pi i \tau_2}
\left( l^n F_{mn}(x+l\tau_2)
 + y_\beta D_{m} \Phi^\beta(x+l\tau_2) \right)
\right. \\
- \int_0^1 d\tau_1~ e^{2\pi i \tau_1} \Phi^\alpha(x+l\tau_1)
\qquad \qquad \qquad \\
\qquad \qquad \times \left. \left.
i \int_0^1 d\tau_2~ e^{-2\pi i \tau_2}
\left( l^n F_{mn}(x+l\tau_2)
+ y_\beta D_{m} \Phi^\beta(x+l\tau_2) \right)\right\}\right].
\end{multline}
\begin{multline}\label{boseab}
T^{\alpha \beta}
= -(2\pi \alpha')^2 \\ \times
\int dx~* \left[e^{ikx}
\exp \left( i\int_0^1 d\tau~ l^i
A_i(x+l\tau) + y_\alpha \Phi^\alpha(x+l \tau) \right)
\right. \\ \times \left.
\left\{
\int_0^1 d\tau_1~ e^{-2\pi i \tau_1}  \Phi^\alpha(x+l\tau_1)
\int_0^1 d\tau_2~ e^{2\pi i \tau_2} \Phi^\beta(x+l\tau_2)
+ (\alpha \leftrightarrow \beta) \right\} \right].
\end{multline}

\newpage


\begin{thebibliography}{99}

\bibitem{Connes:1998cr}
A.~Connes, M.~R.~Douglas and A.~Schwarz,
``Noncommutative geometry and matrix theory: Compactification on tori,''
JHEP{\bf 9802}, 003 (1998)
[hep-th/9711162].

\bibitem{Douglas:1998fm}
M.~R.~Douglas and C.~Hull,
``D-branes and the noncommutative torus,''
JHEP{\bf 9802}, 008 (1998)
[hep-th/9711165].

\bibitem{Cheung:1998nr}
Y.~E.~Cheung and M.~Krogh,
``Noncommutative geometry from 0-branes in a background B-field,''
Nucl.\ Phys.\ B {\bf 528}, 185 (1998)
[hep-th/9803031].

\bibitem{Kawano:1998re}
T.~Kawano and K.~Okuyama,
``Matrix theory on noncommutative torus,''
Phys.\ Lett.\ B {\bf 433}, 29 (1998)
[hep-th/9803044].

\bibitem{Ardalan:1999ce}
F.~Ardalan, H.~Arfaei and M.~M.~Sheikh-Jabbari,
``Noncommutative geometry from strings and branes,''
JHEP{\bf 9902}, 016 (1999)
[hep-th/9810072].

\bibitem{Chu:1999qz}
C.~Chu and P.~Ho,
``Noncommutative open string and D-brane,''
Nucl.\ Phys.\ B {\bf 550}, 151 (1999)
[hep-th/9812219].

\bibitem{Schomerus:1999ug}
V.~Schomerus,
``D-branes and deformation quantization,''
JHEP{\bf 9906}, 030 (1999)
[hep-th/9903205].

\bibitem{Ardalan:2000av}
F.~Ardalan, H.~Arfaei and M.~M.~Sheikh-Jabbari,
``Dirac quantization of open strings and noncommutativity in branes,''
Nucl.\ Phys.\ B {\bf 576}, 578 (2000)
[hep-th/9906161].

\bibitem{Chu:2000gi}
C.~Chu and P.~Ho,
``Constrained quantization of open string in background B field and  
noncommutative D-brane,''
Nucl.\ Phys.\ B {\bf 568}, 447 (2000)
[hep-th/9906192].

\bibitem{Seiberg:1999vs}
N.~Seiberg and E.~Witten,
``String theory and noncommutative geometry,''
JHEP{\bf 9909}, 032 (1999)
[hep-th/9908142].

\bibitem{Yin:1999ba}
Z.~Yin,
``A note on space noncommutativity,''
Phys.\ Lett.\ B {\bf 466}, 234 (1999)
[hep-th/9908152].

\bibitem{Okawa:2000sh}
Y.~Okawa and H.~Ooguri,
``How noncommutative gauge theories couple to gravity,''
hep-th/0012218, to appear in Nucl.\ Phys.\ B.

\bibitem{Ishibashi:2000hs}
N.~Ishibashi, S.~Iso, H.~Kawai and Y.~Kitazawa,
``Wilson loops in noncommutative Yang-Mills,''
Nucl.\ Phys.\ B {\bf 573}, 573 (2000)
[hep-th/9910004].

\bibitem{Das:2000md}
S.~R.~Das and S.~Rey,
``Open Wilson lines in noncommutative gauge theory and tomography of  
holographic dual supergravity,''
Nucl.\ Phys.\ B {\bf 590}, 453 (2000)
[hep-th/0008042].

\bibitem{Gross:2000ba}
D.~J.~Gross, A.~Hashimoto and N.~Itzhaki,
``Observables of non-commutative gauge theories,''
hep-th/0008075.

\bibitem{Mehen:2000vs}
T.~Mehen and M.~B.~Wise,
``Generalized $\ast$-products, Wilson lines and the solution of the  Seiberg-
Witten equations,''
JHEP{\bf 0012}, 008 (2000)
[hep-th/0010204].

\bibitem{Liu:2000mj}
H.~Liu,
``$\ast$-Trek II: $\ast_n$ operations, open Wilson lines
and the Seiberg-Witten map,''
hep-th/0011125.

\bibitem{Das:2001ur}
S.~R.~Das and S.~P.~Trivedi,
``Supergravity couplings to noncommutative branes, open Wilson lines and  
generalized star products,''
JHEP{\bf 0102}, 046 (2001)
[hep-th/0011131].

\bibitem{Hyun:2000ma}
S.~Hyun, Y.~Kiem, S.~Lee and C.~Lee,
``Closed strings interacting with noncommutative D-branes,''
Nucl.\ Phys.\ B {\bf 569}, 262 (2000)
[hep-th/9909059].

\bibitem{Garousi:2000ch}
M.~R.~Garousi,
``Non-commutative world-volume interactions on D-branes and  Dirac-Born-Infeld 
action,''
Nucl.\ Phys.\ B {\bf 579}, 209 (2000)
[hep-th/9909214].

\bibitem{Ambjorn:1999ts}
J.~Ambjorn, Y.~M.~Makeenko, J.~Nishimura and R.~J.~Szabo,
``Finite N matrix models of noncommutative gauge theory,''
JHEP{\bf 9911}, 029 (1999)
[hep-th/9911041].

\bibitem{Ambjorn:2000cs}
J.~Ambjorn, Y.~M.~Makeenko, J.~Nishimura and R.~J.~Szabo,
``Lattice gauge fields and discrete noncommutative Yang-Mills theory,''
JHEP{\bf 0005}, 023 (2000)
[hep-th/0004147].

\bibitem{Dhar:2000wk}
A.~Dhar and S.~R.~Wadia,
``A note on gauge invariant operators in noncommutative gauge theories  and the 
matrix model,''
Phys.\ Lett.\ B {\bf 495}, 413 (2000)
[hep-th/0008144].

\bibitem{Liu:2000ad}
H.~Liu and J.~Michelson,
``$\ast$-TREK: The one loop N = 4 noncommutative SYM action,''
hep-th/0008205.

\bibitem{Dhar:2000nj}
A.~Dhar and Y.~Kitazawa,
``Wilson loops in strongly coupled noncommutative gauge theories,''
hep-th/0010256.

\bibitem{Pernici:2000va}
M.~Pernici, A.~Santambrogio and D.~Zanon,
``The one-loop effective action of noncommutative N = 4 super Yang-Mills  is 
gauge invariant,''
hep-th/0011140.

\bibitem{Garousi:2000ci}
M.~R.~Garousi,
``Transformation of the Dirac-Born-Infeld action
under the  Seiberg-Witten map,''
hep-th/0011147.

\bibitem{Kiem:2000wn}
Y.~Kiem, D.~H.~Park and S.~Lee,
``Factorization and generalized $\ast$-products,''
hep-th/0011233.

\bibitem{Dhar:2001ht}
A.~Dhar and Y.~Kitazawa,
``High energy behavior of Wilson lines,''
JHEP{\bf 0102}, 004 (2001)
[hep-th/0012170].

\bibitem{Liu:2001ps}
H.~Liu and J.~Michelson,
``Supergravity couplings of noncommutative D-branes,''
hep-th/0101016.

\bibitem{Okuyama:2001sw}
K.~Okuyama,
``Comments on open Wilson lines and generalized star products,''
hep-th/0101177.

\bibitem{Gerhold:2000ik}
A.~Gerhold, J.~Grimstrup, H.~Grosse, L.~Popp, M.~Schweda and R.~Wulkenhaar,
``The energy-momentum tensor on noncommutative spaces:
Some pedagogical  comments,''
hep-th/0012112.

\bibitem{Banks:1997vh}
T.~Banks, W.~Fischler, S.~H.~Shenker and L.~Susskind,
``M theory as a matrix model: A conjecture,''
Phys.\ Rev.\ D {\bf 55}, 5112 (1997)
[hep-th/9610043].

\bibitem{Taylor:2001vb}
W.~Taylor,
``M(atrix) theory: Matrix quantum mechanics as a fundamental theory,''
hep-th/0101126.

\bibitem{Douglas:1997yp}
M.~R.~Douglas, D.~Kabat, P.~Pouliot and S.~H.~Shenker,
``D-branes and short distances in string theory,''
Nucl.\ Phys.\ B {\bf 485}, 85 (1997)
[hep-th/9608024].

\bibitem{Sen:1998we}
A.~Sen,
``D0 branes on $T^n$ and matrix theory,''
Adv.\ Theor.\ Math.\ Phys.\ {\bf 2}, 51 (1998)
[hep-th/9709220].

\bibitem{Seiberg:1997ad}
N.~Seiberg,
``Why is the matrix model correct?,''
Phys.\ Rev.\ Lett.\ {\bf 79}, 3577 (1997)
[hep-th/9710009].

\bibitem{Susskind:1997cw}
L.~Susskind,
``Another conjecture about M(atrix) theory,''
hep-th/9704080.

\bibitem{Kabat:1998sa}
D.~Kabat and W.~I.~Taylor,
``Linearized supergravity from matrix theory,''
Phys.\ Lett.\ B {\bf 426}, 297 (1998)
[hep-th/9712185].

\bibitem{Taylor:1999tv}
W.~I.~Taylor and M.~Van Raamsdonk,
``Supergravity currents and linearized interactions for matrix theory  
configurations with fermionic backgrounds,''
JHEP{\bf 9904}, 013 (1999)
[hep-th/9812239].

\bibitem{deWit:1988ig}
B.~de Wit, J.~Hoppe and H.~Nicolai,
``On the quantum mechanics of supermembranes,''
Nucl.\ Phys.\ B {\bf 305}, 545 (1988).

\bibitem{Ishibashi:1997xs}
N.~Ishibashi, H.~Kawai, Y.~Kitazawa and A.~Tsuchiya,
``A large-N reduced model as superstring,''
Nucl.\ Phys.\ B {\bf 498}, 467 (1997)
[hep-th/9612115].


\bibitem{Taylor:1997ik}
W.~I.~Taylor,
``D-brane field theory on compact spaces,''
Phys.\ Lett.\ B {\bf 394}, 283 (1997)
[hep-th/9611042].

\bibitem{Ganor:1997zk}
O.~J.~Ganor, S.~Ramgoolam and W.~I.~Taylor,
``Branes, fluxes and duality in M(atrix)-theory,''
Nucl.\ Phys.\ B {\bf 492}, 191 (1997)
[hep-th/9611202].

\bibitem{Banks:1997nn}
T.~Banks, N.~Seiberg and S.~Shenker,
``Branes from matrices,''
Nucl.\ Phys.\ B {\bf 490}, 91 (1997)
[hep-th/9612157].

\bibitem{Li:1997bi}
M.~Li,
``Strings from IIB matrices,''
Nucl.\ Phys.\ B {\bf 499}, 149 (1997)
[hep-th/9612222].

\bibitem{Ishibashi:1999ni}
N.~Ishibashi,
``p-branes from (p-2)-branes in the bosonic string theory,''
Nucl.\ Phys.\ B {\bf 539}, 107 (1999)
[hep-th/9804163].

\bibitem{Kato:1999mf}
M.~Kato and T.~Kuroki,
``World volume noncommutativity versus target space noncommutativity,''
JHEP{\bf 9903}, 012 (1999)
[hep-th/9902004].

\bibitem{Cornalba:1999hn}
L.~Cornalba and R.~Schiappa,
``Matrix theory star products from the Born-Infeld action,''
hep-th/9907211.

\bibitem{Aoki:2000vr}
H.~Aoki, N.~Ishibashi, S.~Iso, H.~Kawai, Y.~Kitazawa and T.~Tada,
``Noncommutative Yang-Mills in IIB matrix model,''
Nucl.\ Phys.\ B {\bf 565}, 176 (2000)
[hep-th/9908141].

\bibitem{Cornalba:1999ah}
L.~Cornalba,
``D-brane physics and noncommutative Yang-Mills theory,''
hep-th/9909081.

\bibitem{Ishibashi:1999vi}
N.~Ishibashi,
``A relation between commutative and noncommutative descriptions of  D-branes,''
hep-th/9909176.

\bibitem{Okuyama:2000ig}
K.~Okuyama,
``A path integral representation of the map between commutative and  
noncommutative gauge fields,''
JHEP{\bf 0003}, 016 (2000)
[hep-th/9910138].

\bibitem{Bars:2000av}
I.~Bars and D.~Minic,
``Non-commutative geometry on a discrete periodic lattice and gauge  theory,''
Phys.\ Rev.\ D {\bf 62}, 105018 (2000)
[hep-th/9910091].

\bibitem{Kuroki:2000ij}
T.~Kuroki,
``Noncommutativities of D-branes and Theta-changing degrees of freedom in  D-
brane matrix models,''
Phys.\ Lett.\ B {\bf 481}, 97 (2000)
[hep-th/0001011].

\bibitem{Ambjorn:2000nb}
J.~Ambjorn, Y.~M.~Makeenko, J.~Nishimura and R.~J.~Szabo,
``Nonperturbative dynamics of noncommutative gauge theory,''
Phys.\ Lett.\ B {\bf 480}, 399 (2000)
[hep-th/0002158].

\bibitem{Ambjorn:2000bf}
J.~Ambjorn, K.~N.~Anagnostopoulos, W.~Bietenholz, T.~Hotta and J.~Nishimura,
``Large N dynamics of dimensionally reduced 4D SU(N) super Yang-Mills  theory,''
JHEP{\bf 0007}, 013 (2000)
[hep-th/0003208].

\bibitem{Alvarez-Gaume:2000dx}
L.~Alvarez-Gaume and S.~R.~Wadia,
``Gauge theory on a quantum phase space,''
hep-th/0006219.

\bibitem{Seiberg:2000zk}
N.~Seiberg,
``A note on background independence in noncommutative gauge theories,  matrix 
model and tachyon condensation,''
JHEP{\bf 0009}, 003 (2000)
[hep-th/0008013].


\bibitem{Douglas:1996sw}
M.~R.~Douglas and G.~Moore,
``D-branes, Quivers, and ALE Instantons,''
hep-th/9603167.

\bibitem{Douglas:1997pj}
M.~R.~Douglas, H.~Ooguri and S.~H.~Shenker,
``Issues in M(atrix) model compactification,''
Phys.\ Lett.\ B {\bf 402}, 36 (1997)
[hep-th/9702203].

\bibitem{Douglas:1998zw}
M.~R.~Douglas,
``D-branes in curved space,''
Adv.\ Theor.\ Math.\ Phys.\ {\bf 1}, 198 (1998)
[hep-th/9703056].

\bibitem{Douglas:1998ch}
M.~R.~Douglas,
``D-branes and matrix theory in curved space,''
Nucl.\ Phys.\ Proc.\ Suppl.\ {\bf 68}, 381 (1998)
[hep-th/9707228].

\bibitem{Douglas:1998sm}
M.~R.~Douglas, A.~Kato and H.~Ooguri,
``D-brane actions on Kaehler manifolds,''
Adv.\ Theor.\ Math.\ Phys.\ {\bf 1}, 237 (1998)
[hep-th/9708012].

\bibitem{Douglas:1998uy}
M.~R.~Douglas and H.~Ooguri,
``Why matrix theory is hard,''
Phys.\ Lett.\ B {\bf 425}, 71 (1998)
[hep-th/9710178].

\bibitem{Lifschytz:1998pk}
G.~Lifschytz,
``DLCQ-M(atrix) description of string theory, and supergravity,''
Nucl.\ Phys.\ B {\bf 534}, 83 (1998)
[hep-th/9803191].

\bibitem{Taylor:1999gq}
W.~I.~Taylor and M.~Van Raamsdonk,
``Multiple D0-branes in weakly curved backgrounds,''
Nucl.\ Phys.\ B {\bf 558}, 63 (1999)
[hep-th/9904095].

\bibitem{Taylor:2000pr}
W.~I.~Taylor and M.~Van Raamsdonk,
``Multiple Dp-branes in weak background fields,''
Nucl.\ Phys.\ B {\bf 573}, 703 (2000)
[hep-th/9910052].

\bibitem{Myers:1999ps}
R.~C.~Myers,
``Dielectric-branes,''
JHEP{\bf 9912}, 022 (1999)
[hep-th/9910053].

\bibitem{Dasgupta:2000df}
A.~Dasgupta, H.~Nicolai and J.~Plefka,
``Vertex operators for the supermembrane,''
JHEP{\bf 0005}, 007 (2000)
[hep-th/0003280].

\bibitem{Tseytlin:1997cs}
A.~A.~Tseytlin,
``On non-abelian generalisation of the Born-Infeld action in string  theory,''
Nucl.\ Phys.\ B {\bf 501}, 41 (1997)
[hep-th/9701125].

\bibitem{VanRaamsdonk:1999in}
M.~Van Raamsdonk,
``Conservation of supergravity currents from matrix theory,''
Nucl.\ Phys.\ B {\bf 542}, 262 (1999)
[hep-th/9803003].

\end{thebibliography}
\end{document}